\documentclass[aps,prb,10pt,twocolumn,showpacs,floatfix,superscriptaddress,amsmath,amssymb]{revtex4-1}
\usepackage[pdftex]{graphicx}
\usepackage{dcolumn}
\usepackage{bm}
\usepackage{pstricks}
\usepackage{amssymb,amsmath}
\usepackage{pstricks}
\usepackage{gensymb}
\usepackage{soul}
\usepackage{wasysym}

\def\nn{\nonumber}
\newcommand{\ba}{\begin{eqnarray}}
\newcommand{\ea}{\end{eqnarray}}
\def\be{\begin{equation}}
\def\ee{\end{equation}}

\def\bimno{Bi$_3$Mn$_4$O$_{12}$(NO$_3$)}

\begin{document}

\title{Quantum phase diagram of a frustrated antiferromagnet\\ on the bilayer honeycomb lattice}

\author{ Hao Zhang}
\email{zhanghao@iphy.ac.cn}
\affiliation{Beijing National Laboratory for Condensed Matter Physics and Institute of Physics, Chinese Academy of Sciences, Beijing 100190, China}

\author{ Carlos A. Lamas}
\email{lamas@fisica.unlp.edu.ar}
\author{Marcelo Arlego}
\affiliation{IFLP - CONICET, Departamento de F\'isica, Universidad Nacional de La Plata,
C.C.\ 67, 1900 La Plata, Argentina.}

\author{ Wolfram Brenig}
\email{w.brenig@tu-bs.de}
\affiliation{Institute for Theoretical Physics, Technical University Braunschweig, D-38106 Braunschweig, Germany}

%

\begin{abstract}
We study the spin-1/2 Heisenberg antiferromagnet on a bilayer honeycomb
lattice including interlayer frustration. Using a set of complementary
approaches, namely Schwinger bosons, dimer series expansion, bond
operators, and exact diagonalization, we map out the quantum phase
diagram. Analyzing ground state energies and elementary excitation spectra,
we find four distinct phases, corresponding to three collinear magnetic
long range ordered states, and one quantum disordered interlayer dimer
phase. We detail, that the latter phase is adiabatically connected to an
{\em exact} singlet product ground state of the the bilayer which exists
along a line of maximum interlayer frustration. The order within the remaining
three phases will be clarified.\\
\end{abstract}

\maketitle

\section{Introduction}
\label{sec:introduction}	

Disordered phases in frustrated two-dimensional spin systems are a very
active field of research which thrives both, on the synthesis of new
materials as well as the development of new theoretical concepts
\cite{Anderson,ML_2005,Balents_Nature,LM_2011}.  In this context,
Heisenberg antiferromagnets on the honeycomb lattice have attracted
considerable interest recently. \bimno, discovered by Smirnova \textit{et
al.} \cite{BiMnO} is among the materials to display this structure, with
Mn$^{4+}$ ions with $S=3/2$ forming an undistorted honeycomb lattice. Two
honeycomb layers are separated by bismuth atoms, resulting in a bilayer
arrangement, thereby introducing the additional ingredient of a bilayer
honeycomb magnet.

{\it Ab initio} calculations, by Kandpal and Brink \cite{Kandpal} have
resulted in nearest, and frustrating next-nearest neighbor inter- as well
as intralayer exchange as the dominant couplings in \bimno
. Disordered magnetic ground states, which have been observed
experimentally \cite{expnew2}, have been suggested to result from these
competing interactions. While theoretically, substantial progress has been
made regarding the effects of intralayer frustration and quantum disordered
phases in the single-layer honeycomb magnet
\cite{Mulder,Wang,Okumura,Clark,Cabra_honeycomb_prb,Ganesh_2011,Albuquerque,Cabra_honeycomb_2,Mezzacapo,Li_2012_honeyJ1-J2-J3,Bishop_2012,Ganesh_PRL_2013,Zhu_PRL_2013,Zhang_PRB_2013,Fisher_2013,Bishop_2013},
less attention has been given to the the influence of an interlayer
coupling in their impact on disordered phases
\cite{Ganesh_2011,Ganesh_QMC,Oitmaa_2012,Zhang2014}.

The aim of this work is to study the zero temperature phase diagram of a
frustrated Heisenberg model on the bilayer honeycomb lattice including
interlayer frustration. At a particular value of maximum interlayer
frustration we obtain an exactly solvable model, with a dimerised ground
state. We focus on the $S=1/2$ case, where quantum fluctuations become more
important, although some results remain valid for larger values of the
spin, as we discuss in the following. We explore
the quantum phases of the model in the exchange parameter space surrounding
the exact dimer state, using various complementary techniques, including
bond operators (BO), Schwinger boson mean field theory (SB-MFT) and series
expansion (SE) based on the continuous unitary transformation method. These
studies will be complemented with exact diagonalization (ED) using Lanczos
on finite systems. We provide results for ground state energies, spin
gaps, spin correlation functions, the quantum phase diagram, and the nature
of the quantum phase transitions.

The outline of the paper is as follows: Sec. \ref{sec:ground-state}
introduces the model and proves that a product of dimers is the exact
ground state of the system on a special line of the parameter
space. Sec. \ref{sec:QPD} sketches several qualitative aspects of the
the quantum phase diagram. In Sec. \ref{sec:IDP} we analyze the interlayer
dimer phase, departing from the line of the exact dimer state. In
Sec. \ref{sec:Neel} we characterize the magnetic phases, including
N\'eel-like and collinear states. In Sec.  \ref{sec:Quantum-phases} we
summarize our quantitative findings on the quantum phase diagram. In
Sec. \ref{sec:J2} we briefly discuss some consequences of adding intralayer
frustration by next nearest neighbor exchange. Finally in
Sec. \ref{sec:conclusions} we present our conclusions and
perspectives. Several appendices are added for technical details regarding
the methods we use.

\section{model and exact ground state }
\label{sec:ground-state}

\begin{figure}[tb]
\includegraphics[width=0.8\columnwidth]{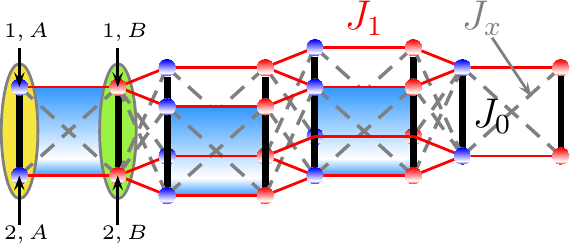}
\caption{(Color online) Dominant exchange interactions in \bimno.  Colored
areas correspond to the unit cells. Frustrating intralayer next
nearest-neighbors interactions are omitted in this figure for simplicity.}
\label{fig:layers}
\end{figure}

We study the Heisenberg Hamiltonian on the bilayer honeycomb lattice
\ba
 \label{eq:Hspin_general}
 H &=&\sum_{\vec{r},\vec{r}',\alpha,\beta}
 J_{\alpha,\beta}(\vec{r},\vec{r}')
 \vec{\bf{S}}_{\alpha}(\vec{r})\cdot \vec{\bf{S}}_{\beta}(\vec{r}'),
\ea
where $\vec{\bf{S}}_{\alpha}(\vec{r})$ is the spin operator on site
$\alpha$ corresponding to the unit cell $\vec{r}$. The index $\alpha$ takes
the values $\alpha=1,A;\; 2,A;\; 1,B;\; 2,B$ corresponding to the four
sites on each unit cell and the couplings $J_{\alpha, \beta} (\vec{r},
\vec{r}')$ are depicted in Fig. \ref{fig:layers}.  As stated in
Sec. \ref{sec:introduction}, the inclusion of the frustrating interlayer
coupling $J_{x}$ is motivated by {\it ab inito} calculations
\cite{Kandpal}. $J_x$ may be comparable to $J_1$ and of relevant magnitude
with respect to the remaining exchange couplings. In Sec. \ref{sec:J2}, we also
consider intralayer next-nearest neighbors frustrated coupling, which will
be labeled $J_2$, but is not shown in Fig. \ref{fig:layers} for simplicity.

In this Section we focus on interlayer frustration only, i.e. $J_2 = 0
$. Interestingly, in that case, the bilayer honeycomb belongs to a class of
Hamiltonians, which exhibits an exact dimer-product ground state in a
certain region of parameter space, even for finite $J_{1,x}$.  This result
is valid for arbitrary site spin $S$.  Hamiltonians with this property
seem to have been constructed first in Ref. \onlinecite{Bose1993a}, based
on methods in Ref. \onlinecite{Bose1992}, and have been reconsidered in
many subsequent studies \cite{Honecker2000a,Brenig2001a,Brenig2002a,Arlego2006a,Lamas2015b}.

\begin{figure}[tb]
\includegraphics[width=0.35\textwidth]{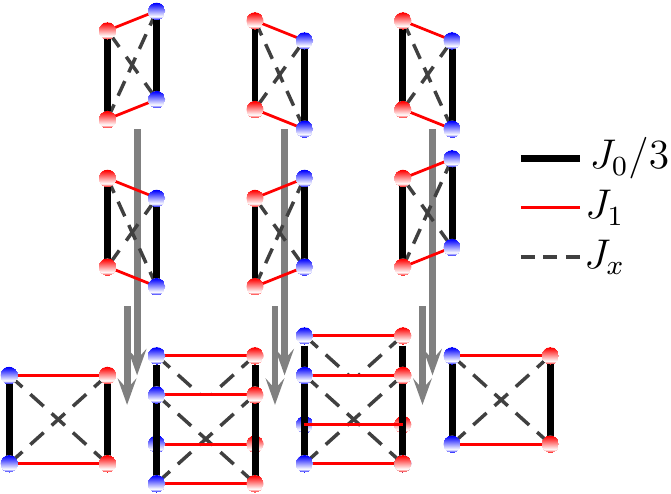}
\caption{(Color online) Decomposition of the Heisenberg model on the
frustrated bilayer honeycomb lattice into three sets of four-spin
plaquets.}
\label{fig:exact_bilayer}
\end{figure}

Using Fig. \ref{fig:exact_bilayer}, we start by writing the Hamiltonian
Eq. (\ref{eq:Hspin_general}) as $H=H_{0}+H_{1}+H_{2}$, with
\begin{eqnarray}10
\lefteqn{
H_{i}=\sum_{\vec{r}}\bigg[
\frac{J_{0}}{3}\left( \vec{\mathbf{S}}_{1,A}(\vec{r_i})\cdot
\vec{\mathbf{S}}_{2,A}(\vec{r_i})
+\vec{\mathbf{S}}_{1,B}(\vec{r})\cdot
\vec{\mathbf{S}}_{2,B}(\vec{r})\right)
\nonumber}
\\
&+& J_{1}\left(
\vec{\mathbf{S}}_{1,A}(\vec{r_i})\cdot \vec{\mathbf{S}}_{1,B}(\vec{r_i})+
\vec{\mathbf{S}}_{2,A}(\vec{r_i})\cdot \vec{\mathbf{S}}_{2,B}(\vec{r_i})
\right)\hphantom{aaaaaa}\nonumber\\
&+&  J_{x}\left(
\vec{\mathbf{S}}_{1,A}(\vec{r_i})\cdot \vec{\mathbf{S}}_{2,B}(\vec{r_i})+
\vec{\mathbf{S}}_{2,A}(\vec{r_i})\cdot \vec{\mathbf{S}}_{1,B}(\vec{r_i})
\right)\bigg],
\end{eqnarray}
in which $i=0,1,2$ corresponds to $\vec{r}_{ (0,1,2) }= \vec{r} +(\vec{0}
,\vec{e}_{1} ,\vec{e}_{2})$, being $\vec{e}_{1}$ and $\vec{e}_{2}$ the
primitive vectors of the triangular lattice. Introducing the
bond spin operators
\begin{equation}
\label{eq:L-K}
\vec{\bf L}_{\alpha}=\vec{\mathbf{S}}_{1,\alpha}+\vec{\mathbf{S}}_{2,\alpha}
\hspace{1cm}
\vec{\bf K}_{\alpha}=\vec{\mathbf{S}}_{1,\alpha}-\vec{\mathbf{S}}_{2,\alpha}.
\end{equation}
with $\alpha=A,B$, we can rewrite $H_0$ as
\ba \nonumber
\label{eq:H0}
H_{0}&=&-2J_{0}NS(S+1)+\sum_{\vec{r}}\left\{ \frac{J_{0}}{2}\left(
\vec{\mathbf{L}}_{A}^{2}(\vec{r})
+\vec{\mathbf{L}}_{B}^{2}(\vec{r})\right)\right.\\\nonumber
&+&
\left( \frac{J_{1}+J_{x}}{2}\right)\left( \vec{\mathbf{L}}_{A}(\vec{r})
\cdot\vec{\mathbf{L}}_{B}(\vec{r})\right)\\
&+&\left.
\left( \frac{J_{1}-J_{x}}{2}\right)\left(
\vec{\mathbf{K}}_{A}(\vec{r}) \cdot\vec{\mathbf{K}}_{B}(\vec{r})\right)
\right\},
\ea
with similar expressions for  $H_{1}$ and $H_{2}$.

The main point of this Section is, that for $J_{1}=J_{x}$, the last term in
the Hamiltonian vanishes, and therefore, (i) {\em each} bond spin
$\vec{\mathbf{L}}_{A}(\vec{r})$ is conserved and (ii) the {\em total} bond
spin $\sum_{\vec{r}} \vec{\mathbf{L}}_{A} (\vec{r})$ is
conserved. Therefore, at $J_{1}=J_{x}$, the eigenstates of $H$ are
multiplets of the total bond spin. Among those is the {\em product state of
bond singlets}, i.e. $|\psi\rangle = \bigotimes_{ i=1 }^{N} | s_{A}
\rangle_{ \vec{r}_i} |s_{B} \rangle_{\vec{r}_i}$ with $\vec{\bf
L}_{\alpha}(\vec{r}_i)|s_{\alpha}\rangle_{\vec{r}_i}=0$, and $|s_\alpha
\rangle_{\vec{r}_i} = \sum_{m =-S}^S (-1)^{S-m} |m,-m\rangle /\sqrt{2
S+1}$. Here $|m,-m\rangle$ labels a product of eigenstates of ${\bf
S}^{z}_{1;\alpha}(\vec{r}_i)$ and ${\bf S}^{z}_{2;\alpha}(\vec{r}_i)$ on
dimer $\alpha$ of the unit cell located at $\vec{r}_{i}$. The energy
$E_{0}$ of $|\psi \rangle$ can be read off from Eq. (\ref{eq:H0}), namely
$E_{0}=-J_0 N S(S+1)$.

For any other multiplets of the {\em total} bond spin one has to promote
dimers into eigenstates of $\vec{\mathbf{L}}_\alpha(\vec{r})$ different
from zero. This will increase any eigenstate's energy proportional to
$J_0$, due to the first term under sum in Eq. (\ref{eq:H0}), but will also
lead to exchange-lowering of the energy proportional to $J_1+J_x$ from
pairs of nearest neighbor dimers with non-zero bond spin due to the second
term under sum in Eq. (\ref{eq:H0}). Therefore, for any finite site spin
$S$, and for $J_1$ less than a critical coupling $0<J_1<J^c_1$, $|\psi\rangle$
is indeed also the ground state at $J_{1}=J_{x}$.

While we emphasize, that the preceding is valid for {\em any} site spin
$S$, the nature of the state for $J_1>J^c_1$ at $J_{1}=J_{x}$ may depend on
details. However, for $S=1/2$ the situation is definite. Since there are
only two eigenstates of $\vec{\mathbf{L}}_{A}(\vec{r})$, i.e. singlet and
triplet, the ground state will either be $|\psi\rangle$ or stem from the
sector of {\em all} $\vec{\mathbf{L}}_{\alpha} (\vec{r})$ in triplet states
$|t_{\mu\alpha} \rangle_{\vec{r}_i}$, where $\mu$ refers to the
z-component. By virtue of Eq. (\ref{eq:H0}) the latter sector is isomorphic
to the spin-1 Heisenberg model on the hexagonal lattice. In both of these
sector nucleation of inhomogeneous distributions of $L=0$ and $L=1$ are
energetically unfavorable, i.e. do not lead to ground states.

The exact dimer singlet product state serves as a convenient starting point
for several perturbative and mean field methods, which we will take
advantage of starting with Sec. \ref{sec:IDP}.

\section{Qualitative aspects}
\label{sec:QPD}

\begin{figure}[tb]
\includegraphics[width=0.45\textwidth]{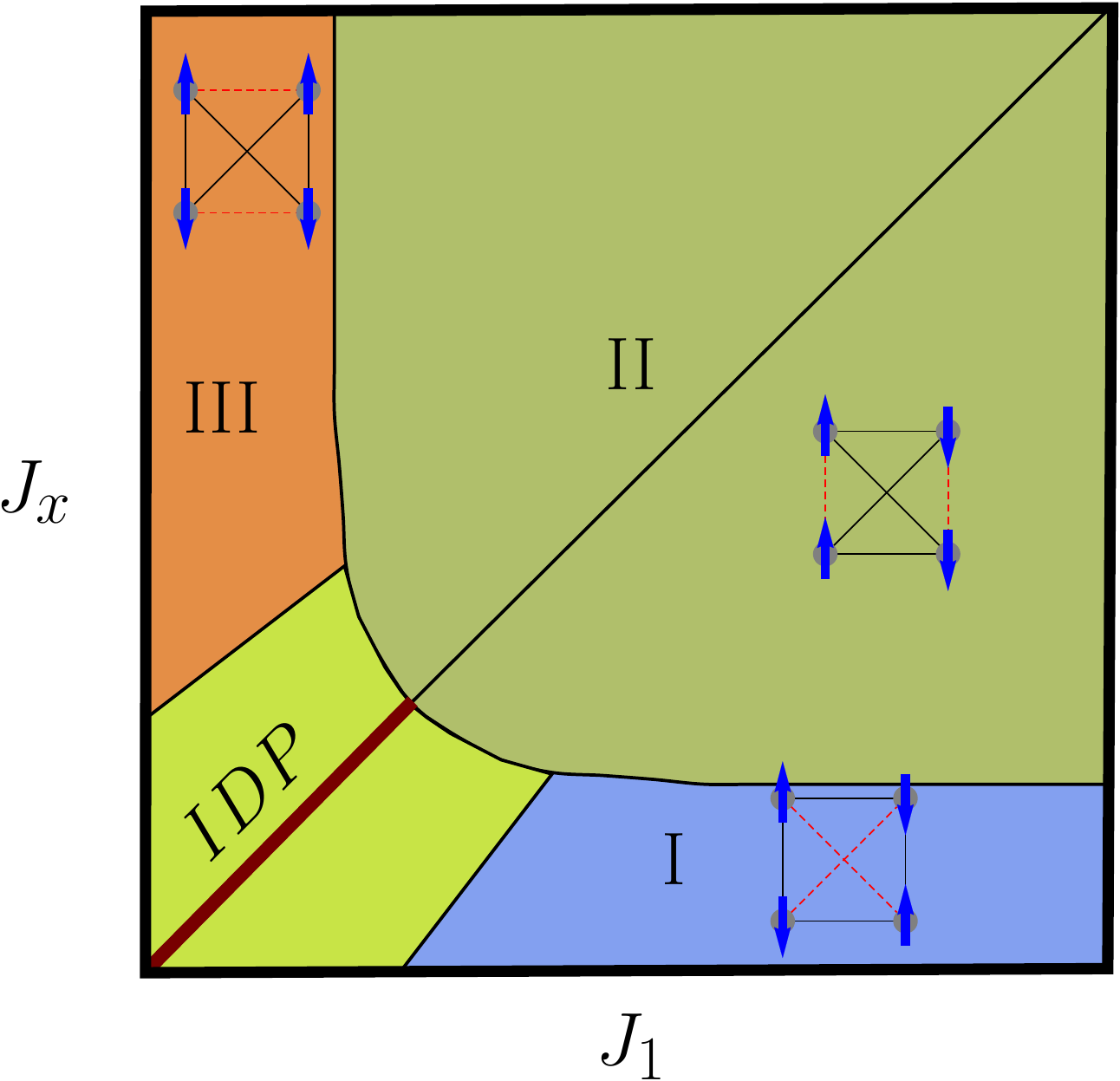}
\caption{Qualitative sketch of quantum phase diagram of Heisenberg model on
the frustrated bilayer honeycomb lattice}
\label{fig:phase-diagram}
\end{figure}

In order to pave the way through the remainder of this work, we provide a
{\it qualitative} picture of the quantum phase diagram to be expected for the
bilayer without intralayer frustration ($J_2=0$) in this Section. This is
depicted in Fig. \ref{fig:phase-diagram}. A {\it quantitative} justification will
be given in the following Sections by analyzing various regions of this
anticipated phase diagram, considering ground state energies, low energy
excitations, triplet gaps, order parameters and spin correlations as
extracted from complementary methods, specifically exact diagonalization,
Schwinger boson and bond operator mean field theories, series expansion and
linear spin-wave theory.

Several comments apply to Fig. \ref{fig:phase-diagram}. First, the diagram
is symmetric respect to the $J_{1}=J_{x}$ line. This is evident at the
Hamiltonian level. Indeed, from Fig. \ref{fig:layers} we see that
exchanging $J_{1} \leftrightarrow J_{x}$, induces a site
exchange $1,B \leftrightarrow 2,B$, which in turn results in $K_B
\leftrightarrow -K_B$. This leaves the last term of $H_0$ in
Eq.(\ref{eq:H0}) invariant. The same is true for $H_1$ and $H_2$.  In the
following we normalize energies in units of $J_0$ and introduce the
dimensionless couplings $j_0=1$, $j_{1}=J_{1}/J_{0}$, $j_{2}=J_{2}/J_{0}$
and $j_{x}=J_{x}/J_{0}$.

The bold dark-red section of the diagonal line of maximum frustration,
$j_{1}=j_{x}$ in Fig. \ref{fig:phase-diagram}, refers to the exact dimer
state. As discussed in Sec. \ref{sec:ground-state} this state terminates in
a first order transition point into the ground state of an $S=1$ AFM
Heisenberg on the single layer hexagonal lattice, which extends over the
solid black diagonal line shown in Fig. \ref{fig:phase-diagram}. We will
show, that this occurs at $j_{1}=j_{x}\simeq 0.5$.

Departing off the line of maximum frustration the exact dimer turns into a
gaped \emph{interlayer dimer phase} (IDP) (see Fig. \ref{fig:phase-diagram}).
This phase is quantum disordered, and shows
dispersive triplon excitations. The triplon gap will decrease from $\Delta=1$
as distance increases from the diagonal line.

For sufficiently large $j_1$ and/or $j_x$, the system will favor
collinear order with a straightforward semiclassical interpretation. Namely
three possibilities exist to minimize two out of the three exchange energies,
leaving one of them frustrated. The corresponding spin arrangements and
phases are labeled I, II, and III in
Fig. \ref{fig:phase-diagram}, with the frustrated link marked by red dashes.
Phases I and III obey the $j_{1} \leftrightarrow j_{x}$ symmetry already
mentioned. While the classical states I, II and III do not represent
exact eigenstates of the Hamiltonian, we detect signals of these orderings
in the quantum model, which justify this identification.

We end this Section by expressing some expectations, regarding the order of the phase
transitions. Since the symmetry of phases I, II, and III have no subgroup
relations, we expect the transitions I-II and II-III to be of first order,
i.e. of level-crossing type. On the other hand, the transition from the IDP into the magnetic
phases I and III will be signaled by the closure of the IDP spin gap
$\Delta$, which decreases symmetrically from 1 to 0, off the red
exact-dimer product line up to the two corresponding critical lines. This gap closure
signals a second order quantum phase transition.

Finally, as discussed in Sec. \ref{sec:ground-state}, the transition from the tip of
the bold dark-red line in the IDP to phase II is first order. The nature of the
transition remains first order all along the IDP-II transition up to the two
tricritical points, separating IDP-I-II and IDP-II-III phases.

\section{Interlayer Dimer Phase}
\label{sec:IDP}

In this Section we analyze the interlayer dimer phase (IDP) at
$j_{1},j_{x}\ll1$.   In particular, we discuss our results for
the ground state energy and the spin gap, as obtained from dimer series
expansion (D-SE), bond operator (BO) theory using Holstein-Primakoff (HP)
and mean-field theory (MFT), as well as from exact diagonalization
(ED). Both, D-SE and BO-HP/MFT are natural approaches to treat the IDP,
since they are both exact in the fully decoupled dimer-product state, along
the line $j_{1}=j_{x}$ and treat deviations from the latter
perturbatively. While D-SE is exact order-by-order in
$j_{1}-j_{x}$, BO-HP/MFT is perturbatively proper only
to leading order.  Since both approaches renormalize only the fully
decoupled dimer-product state, they are insensitive to level crossing,
which may occur within the ground state, as a function of
$j_{1}-j_{x}$. In turn, these methods do not detect first
order, but only second order quantum phase transitions accompanied by the
closure of a spin gap. Therefore, in order to probe for first order
transitions, we resort to ED as an unbiased technique. While finite size
effects, render ED less effective to detect gap closures, it allows to
search for level crossings rather effectively. In turn ED, BO, and D-SE are
complementary to determine the extent of IDP phase, as well as the nature
of the transitions also to the other phases present in the model. Technical
details about the implementation of the different methods can be found in the
Appendices.

\begin{figure}[tb]
\centering{}\includegraphics[width=0.75\columnwidth]{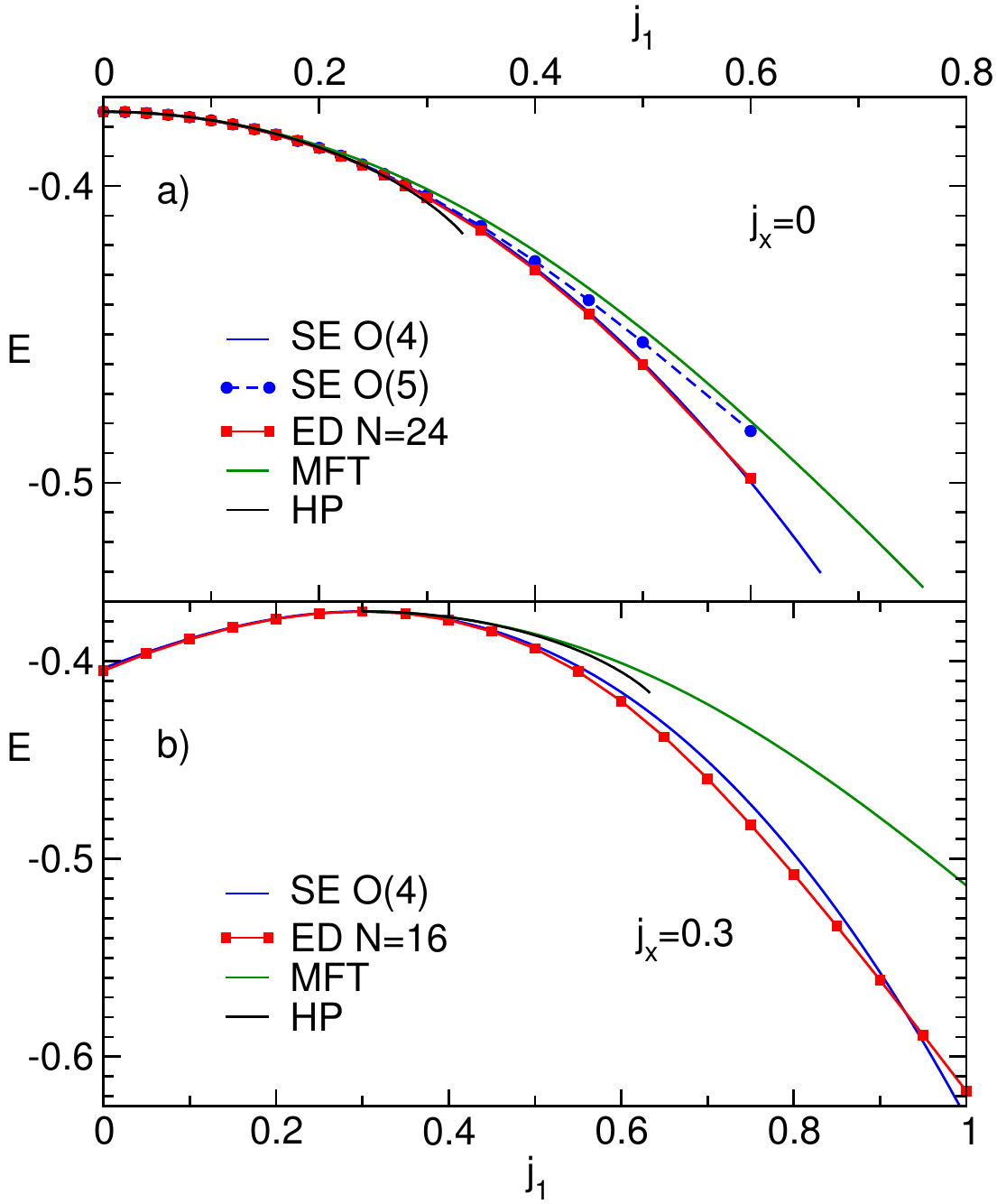}
\protect\caption{(Color online) Ground state energy per spin $E$ at $j_{2}=0$
versus $j_{1}$ from ED (red with squares), D-SE (blue, blue dashed with
circles), BO-HP (black) and BO-MFT (green) for a) $j_{x}=0$ with system
size $N=24$, and orders $O(4)$ and $O(5)$, see also
Refs. \onlinecite{Zhang2014,GE1,Arlego201415} and b) $j_{x}=0.3$ with system size
$N=16$, and order $O(4)$.}
\label{fig:eg-bilayer}
\end{figure}

We begin by considering the ground state energy. From D-SE we obtain
the following O(4) expression for the ground state energy \emph{per
spin} evolving from the limit of decoupled interlayer dimers
\begin{eqnarray}
E(j_{1},j_{x}) & = & -\frac{3}{8}+\frac{9}{512}\left(j_{1}-j_{x}
\right)^{2}\Big[-16-8(j_{1}+j_{x})\nonumber \\
 &  & +3\left(j_{1}^{2}+j_{x}^{2}\right)-22j_{1}j_{x}\Big]
\,.\label{eq:egSE}
\end{eqnarray}
This explicitly satisfies $E(j_{1},j_{1})=-\frac{3}{8}$, corresponding the
exact dimer-product solution along $j_{x}=j_{1}$ and
$E(j_{1},j_{x})=E(j_{x},j_{1})$ fulfilling the
Hamiltonian invariance under $j_{1}\leftrightarrow j_{x}$.  In
Fig. \ref{fig:eg-bilayer} we compare the ground state energy obtained from
the various methods for two different values of $j_{x}$.
Fig. \ref{fig:eg-bilayer}a), in part also contains BO-MFT solutions from
Refs. \onlinecite{GE1,Arlego201415} and results from Ref. \onlinecite{Zhang2014}, where $O(5)$ D-SE is
available at $j_{x}=0$, and ED for $N=24$ sites.  In both panels
and for all methods, the energy shows a maximum at $j_{1}=j_{x}$,
where the ground state is a dimer-product state with energy per spin equal
to $-3/8$. Around the exact solution point, ED and D-SE show excellent
agreement up to $|j_{1}-j_{x}|\simeq0.2\ldots0.3$ in both
panels. Deviations between ED and D-SE beyond that points are due to finite
size effects of the ED and due the finite order of the D-SE. The impact of
the latter can be assessed at $j_{x}=0$, where higher orders of the
D-SE have been reached \cite{Zhang2014}.  From Fig. \ref{fig:eg-bilayer}a),
clearly visible differences arise between $O(4)$ and $O(5)$ D-SE for
$|j_{1}-j_{x}|\gtrsim0.3$.  Turning to the BO theory, two comments
are in order. First, the HP spin gap closes within the range of
$j_{1}$, $j_{x}$-values depicted.  Therefore, the BO curves
terminate. Second, both HP and MFT depend on $j_{1}$ and $j_{x}$
only via the difference $j_{1}-j_{x}$. This is not an exact
property of the model beyond leading order, which is obvious e.g. from
Eq. (\ref{eq:egSE}). In turn, BO results are identical for
Fig. \ref{fig:eg-bilayer}a) and b) up to a shift of origin and have been
plotted only for positive $j_{1}-j_{x}$.  Moreover, agreement
between ED, D-SE and BO is expected to be best at either $j_{1}=0$ or
$j_{x}=0$, which is consistent with this figure. In fact, the
agreement between all four methods is excellent for $j_{x}=0$ and
for $j_{1}\lesssim0.3$, while ED and D-SE show some difference to BO theory
at $j_{x}=0.3$. In view of the significant changes from $O(4)$ to
$O(5)$ D-SE, a quantitative assessment of these differences is beyond this
work. In fact, Fig.  \ref{fig:eg-bilayer}a) would suggest that $O(5)$ D-SE
agrees better with BO theory than with ED for $j_{1}\gtrsim0.3$.

\begin{figure}[tb]
\centering{}\includegraphics[width=0.9\columnwidth]{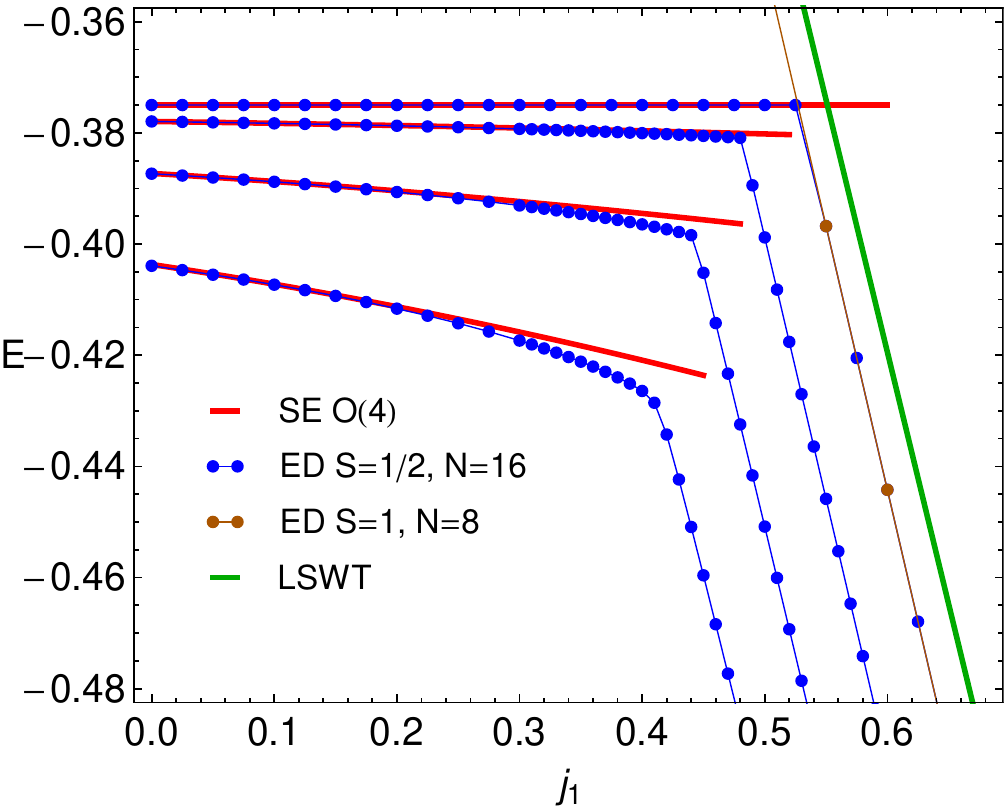}
\protect\caption{(Color online) Ground state energy per spin $E$ vs $j_1$, for different
paths parametrized by $b=j_{x}-j_{1}$, with $b=$ 0, 0.1, 0.2, and 0.3
(top to bottom). Line-connected blue (brown) dots: ED for S=1/2 (S=1)
bilayer (effective single layer). Solid red: D-SE. Green: LSWT for S=1
effective single layer.}
\label{fig:energy-exact}
\end{figure}

\begin{figure}[t]
\centering{}\includegraphics[width=0.75\columnwidth]{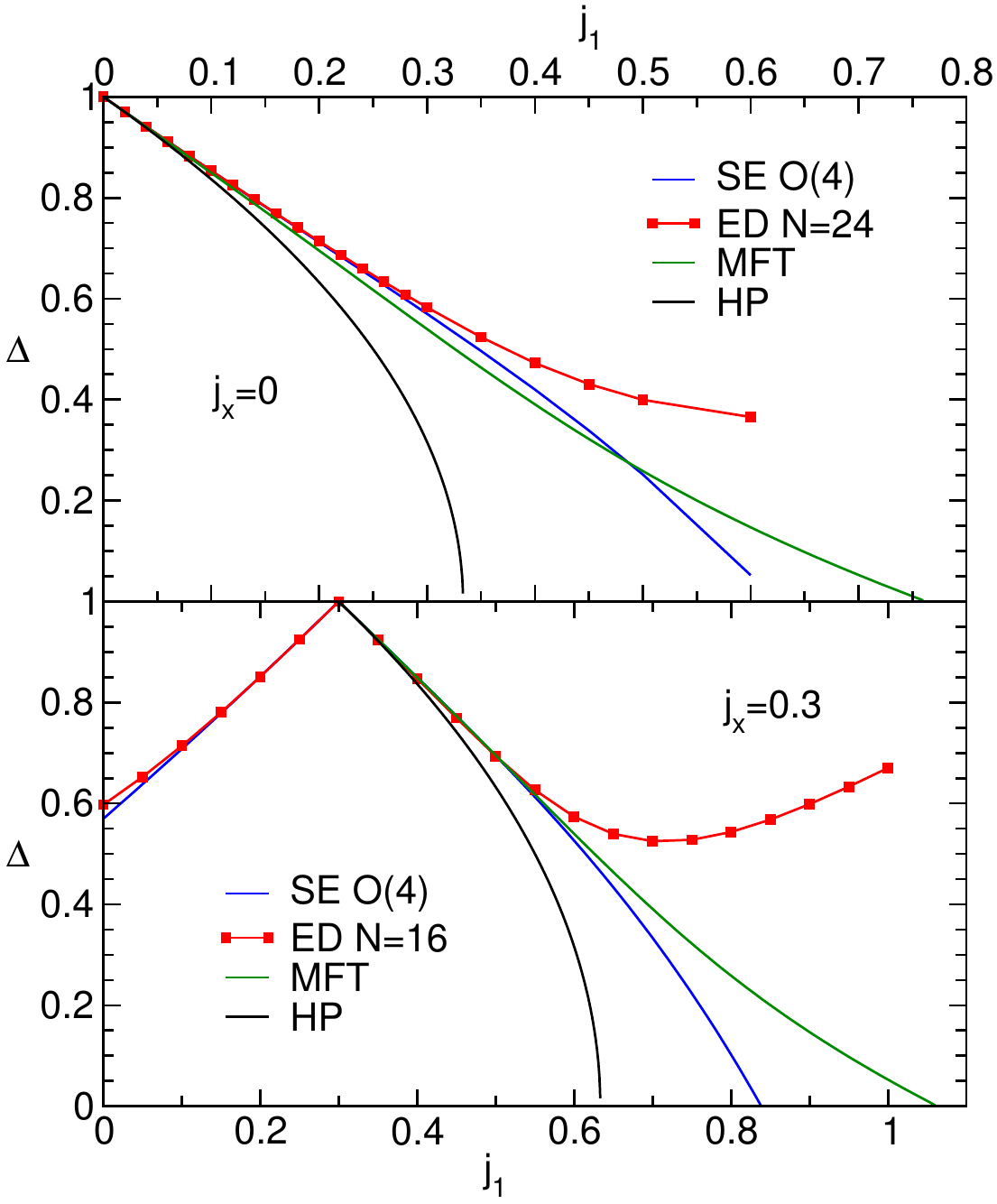}
\protect\caption{(Color online) Spin gap $\Delta$
versus $j_{1}$ from ED (red with squares), $O(4)$ D-SE (blue), BO-HP
(black), and BO-MFT (green), for a) at $j_{x}=0$ with system size
$N=24$, see also refs. \cite{Zhang2014,GE1,Arlego201415} and b) $j_{x}=0.3$
with system size $N=16$.}
\label{fig:gap-bilayer}
\end{figure}

While the variations of results between the methods discussed so far are
quantitative only, we expect a qualitative difference between ED and D-SE
or BO theory in the vicinity of the first order transition from the IDP to
the magnetic phase II (Fig. \ref{fig:phase-diagram}). Therefore, in
Fig. \ref{fig:energy-exact} we depict the ground state energy per spin versus $j_{x}$
along lines parametrized by $b=j_{x}-j_{1}$, with $b=$ 0, 0.1, 0.2, and 0.3
from top to bottom. ED results are shown by line-connected blue dots,
whereas D-SE results are shown by solid red lines. First, the small, albeit
finite slope of $E$ at small $j_{1}$ in this figure, which is increasing as
$b$ increases, demonstrates once more, that properties of the system in the
IDP are not only functions of $b=j_{x}-j_{1}$.  Therefore, in this figure
we do not consider BO results. Second, we note that for $b=0$ ($j_{1} =
j_{x}$) the upper pair of curves representing ED and D-SE coincide exactly
at $-3/8$ up to a critical point of $j_{1}^{c} , j_{x}^{c} \simeq
0.52$. This corresponds to the bold red line in
Fig. \ref{fig:phase-diagram}.  At the critical point, ED exhibits a kink in
the energy versus $j_{1}$, signaling a first order transition into another
type of ground state of the system. Clearly D-SE cannot detect this
transition because it adiabatically evolves the dimer state with $j_{1}$,
which discontinues to be the ground state for $j_{1} >
j_{c}^{c}$. Qualitative differences between ED and D-SE are also observed
off the diagonal line, for $j_{1}$ roughly larger than $j_{1}^{c}$.  Here
again, a clear change of slope is detected by ED in
Fig. \ref{fig:energy-exact} for $b = 0.1 , 0.2$. This supports our claim
that the transition IDP-II is first order, as anticipated in the previous
Section. At $b = 0.3$, ED shows no clear signature of a single kink
anymore, suggesting a succession of second and then first-order
transitions, close to one of the tricritical points of
Fig. \ref{fig:phase-diagram}.

Non-IDP phases will be analyzed in detail in the following Sections. Here
we elaborate further on the transition from the IDP into the effective
$S=1$ AFM on the single layer hexagonal lattice anticipated already in
Sec. \ref{sec:ground-state}. We have verified this scenario using two
checks. First, we have performed ED calculations on a \emph{single} layer
spin-1 cluster comprising the same \emph{site}-geometry as that of the
\emph{dimers} in the original cluster. The corresponding ground state
energy is depicted by line-connected brown dots in
Fig. \ref{fig:energy-exact}.  The excellent agreement between both types of
ED calculations verifies our assertion. For a second check, we have
considered linear spin wave theory (LSWT) for the ground state energy of
the spin-1 Heisenberg antiferromagnet on the hexagonal lattice. For details
see appendix \ref{sec:Linear-Spin-Wave}. The result, also shown in
Fig. \ref{fig:energy-exact}, is quantitatively very similar to the ED
results, with $j_{1}^{c}\simeq0.551$.  Since LSWT for a collinear state
with $S=1$ should be rather well defined, it would be interesting to
analyze if the small difference of the critical coupling $\Delta
j_{1}^{c}\approx0.03$ between ED and LSWT is dominated by $O(1/S^{2})$
correction or by finite size effects.

Perpendicular to the exact dimer line, the dispersion of triplons will lead
to a closure of the spin gap $\Delta$ for sufficiently large $j_{1}-j_{x}$.
From $O(4)$ D-SE we get
\begin{eqnarray}
\Delta(j_{1},j_{x}) & = & 1-\frac{3}{16}\left|j_{1}-j_{x}\right|\left|-8+\left(j_{1}-j_{x}\right)^{2}(j_{1}-
j_{x})\right|\nonumber \\
 &  & -\frac{3}{128}\left(j_{1}-j_{x}\right)^{2}\Big[-16+
8(j_{1}-j_{x})\nonumber \\
 &  & +55\left(j_{1}^{2}+j_{x}\right)-14j_{1}j_{x}\Big]
\,.\label{eq:gapSE}
\end{eqnarray}
As for the ground state energy, Eq. (\ref{eq:egSE}), this satisfies
$\Delta(j_{1},j_{x})=\Delta(j_{x},j_{1})$ and resembles the decoupled dimer
state, i.e. $\Delta(j_{1},j_{1})=1$. In Fig. \ref{fig:gap-bilayer} we
compare Eq. (\ref{eq:gapSE}) with ED, BO-HP and BO-MFT versus $j_{1}$ for
the same two values of $j_{x}$ as in Fig. \ref{fig:eg-bilayer}. As for the
ground state energy, the BO results are identical for
Fig. \ref{fig:gap-bilayer}a) and b) up to a shift of origin and have been
plotted only for positive $j_{1}-j_{x}$. As is clear from the figure, ED,
D-SE, and BO-MFT tend to keep the spin gap open for a larger range of
exchange couplings off the exact dimer state, while the BO-HP gap closes
more rapidly. The agreement between ED, D-SE, and BO-MFT is very good for
$|j_{1}-j_{x}|\lesssim0.3$. It is obvious, that finite size effects for the
spin gap in the ED are rather large, showing a minimum of $\Delta$ of
$\sim0.35$ at $N=24$, versus $\sim0.5$ only for $N=16$. A proper
finite-size scaling analysis of the spin gap from ED is unfeasible, because
of the large unit cell. Interestingly, while BO-HP shows standard square
root behavior of the gap at the critical point, with a negative curvature,
self-consistency within the BO-MFT leads to a positive curvature of
$\Delta$, with no obvious power law at gap closure.

\begin{figure}[t]
\includegraphics[width=0.45\textwidth]{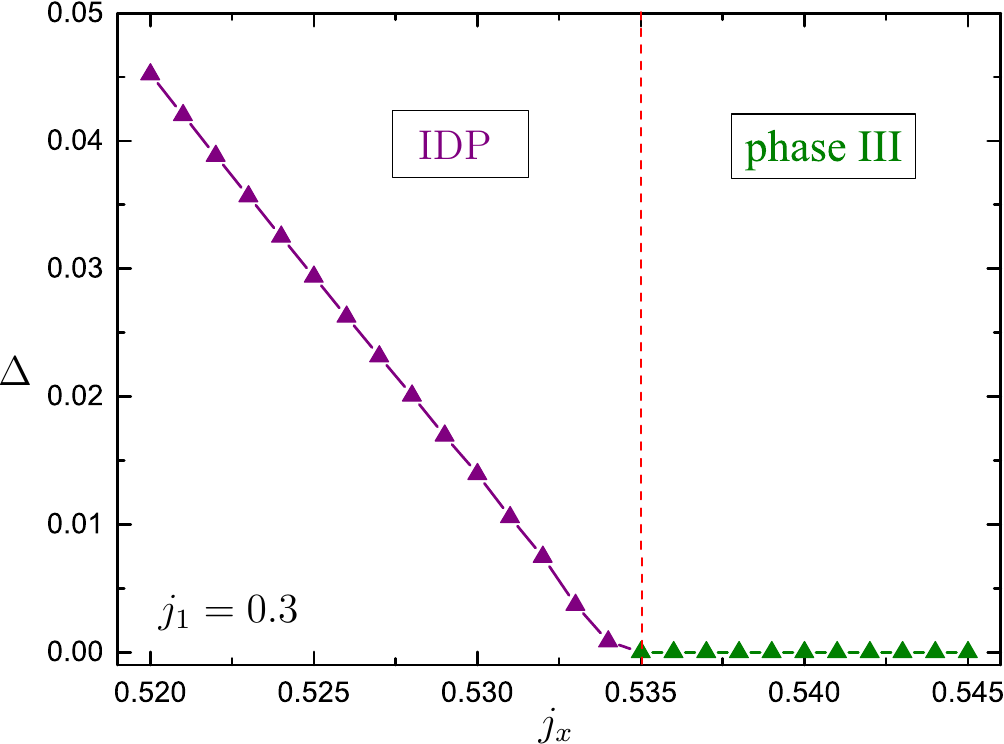}
\caption{(Color online) Example of SB-MFT gap along vertical cut
through the phase diagram Fig. \ref{fig:phase-diagram}, at $j_1=0.3$ for
the IDP-III transition and extrapolated to the thermodynamic limit.}
\label{fig:continuous_transition}
\end{figure}

We close this Section with two remarks on SB-MFT. Also in this approach,
quantum disordered phases are associated with a gapped excitation
spectrum. In turn, the IDP can equally well be detected using SB-MFT.
However, while in the D-SE and BO theory the elementary excitations in the
IDP actually correspond to the physical triplons, in SB-MFT they are
fractionalized bosonic spinons. The latter are unphysical in the IDP. In
order to obtain a proper spin spectrum and the gap, the two-spinon
propagator would have to be evaluated, see e.g. Ref. \onlinecite{Mezio1997a}, however
{\em including} interactions beyond Ref. \onlinecite{Mezio1997a}, in order to confine
the spinon into a sharp triplon mode.  We will not perform such
calculations. Despite of this, it is perfectly valid to use SB-MFT to
extract transition points from the IDP into the magnetic phases of the
bilayer from a closure of the spinon gap, since long range magnetic order
is characterized by a condensation of the bosons at some wave vector
leading to a gapless spectrum. In Fig. \ref{fig:continuous_transition} we
show a representative example. As the second remark, let us note that
SB-MFT predicts a critical point $j_{1}^{c}=0.547$ on the $j_{1}=j_{x}$
line for the transition IDP-II, which agrees very well with the LSWT
prediction given by $j_{1}^{c}=0.551$, and therefore is larger than ED,
similar to the latter.

\section{Magnetic phases}
\label{sec:Neel}

\begin{figure}[tb]
\includegraphics[width=0.48\textwidth]{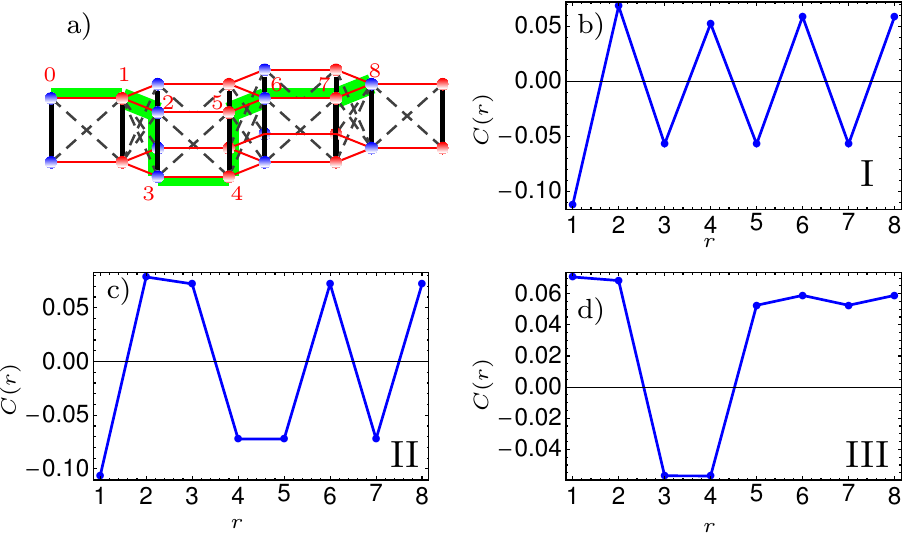}
\caption{(Color online) Static correlation function $C(r)$ vs. $r$ along
the green path depicted in panel a), obtained by means of ED on a finite
cluster of 16 spins. Panel b) $j_{1,x}=0.7,0.3$, c) $j_{1,x}=0.7,0.7$, and
d) $j_{1,x}=0.3,0.7$ clearly show a pattern consistent with the classical
structure shown in regions I, II, and III of Fig. \ref{fig:phase-diagram}.}
\label{fig:corr-ED}
\end{figure}

In this Section we analyze quantum properties of the phases I, II and III of
Fig. \ref{fig:phase-diagram}.  These are gapless states with magnetic long-range order (LRO)
and a spin structure, which has been explained on the classical level in Sec.
\ref{sec:QPD}.

To investigate how the signatures of these orderings survive under quantum
fluctuations, we evaluate the static correlation functions $C(r)=\langle
S_0^z S_r^z \rangle $. In panels (b-d) of Fig. \ref{fig:corr-ED} we show
$C(r)$ vs $r$ along the green path depicted in panel (a), calculated by
means of ED on a finite cluster of 16 spins. We have selected three
different points of parameters space to illustrate the behavior of the
correlations along the considered path. In panel (b) we show $C(r)$ for the
point $(j_{1}=0.7, j_{x}=0.3)$, whereas in panel (d) we depict the
correlation for the symmetric point $(j_{x}=0.7, j_{1}=0.3)$. As it can be
observed, in both cases the sign alternation in $C(r)$ is consistent with
the magnetically ordered phases I and III illustrated in the insets of
Fig. \ref{fig:phase-diagram}.  The same occurs with panel (c), which shows
$C(r)$'s dependence on $r$ for $(j_{x}=0.7, j_{1}=0.7)$.  In this case the
behavior of the correlation is consistent with the classical spin pattern
depicted in the inset of phase II in Fig. \ref{fig:phase-diagram}.

Although we can verify short-distance correlations consistent with the
ordered phases by means of ED, the finite cluster size imposes severe
constraints and does, for example, not permit to consider the actual form of
$C(r)$ and to claim LRO in the sense of $C(r{\rightarrow}\infty){=} const.$
These aspects can be considered with complementary techniques, such as
Schwinger bosons mean field theory (SB-MFT). This approach has been
successfully used to study two-dimensional frustrated Heisenberg
antiferromagnets
\cite{Sachdev,Zhang_PRB_2013,Zhang2014,Cabra_honeycomb_prb,Trumper2}.  We
refer to Appendix \ref{sub:SW-MFT} for details about this technique.

\begin{figure}[t!]
\includegraphics[width=0.45\textwidth]{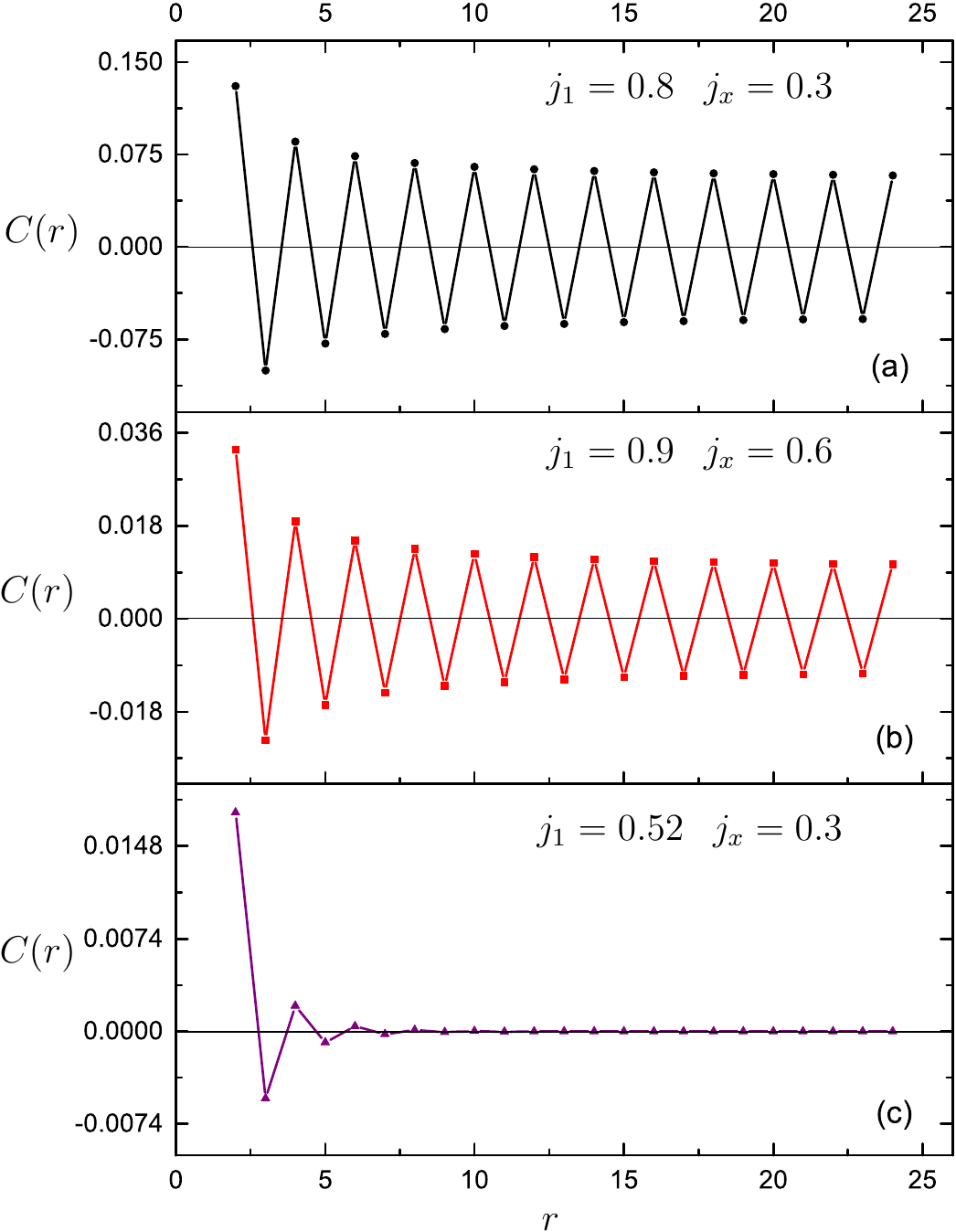}
\caption{ (Color online) Spin-spin correlation between spins belonging to
the same layer in the zigzag direction obtained by SBMFT for a 10000 sites
system. It is shown for the three different phases in the $j_{1}>j_{x}$
side of the phase diagram (Fig. \ref{fig:phase-diagram}): (a) $j_{1}=0.8$,
$j_{x}=0.3$ (phase I), (b) $j_{1}=0.9$, $j_{x}=0.6$ (phase II), and (c)
$j_1=0.52$, $j_x=0.3$ (IDP).}
\label{fig:SSCF_zigzag}
\end{figure}

Fig. \ref{fig:SSCF_zigzag} shows the spin-spin correlation calculated by
means SB-MFT between spins belonging to the same layer, and traversing the
layer along one of the 'zigzag-chain' paths of the hexagonal lattice, for a
system of 10000 sites at $j_{1}=0.8$, $j_{x}=0.3$ (phase I); $j_{1}=0.9$,
$j_{x}=0.6$ (phase II); and $j_{1}=0.52$, $j_{x}=0.3$ (IDP). The last case
is depicted for a contrast to the magnetic phases. Due to the mirror
symmetry of the phase diagram along the line $j_{1}=j_{x}$, we confine the
figure to $j_1 \geqslant j_x$. While AFM LRO is clearly visible in
panels (a) and (b) on each layer, the difference between (a) and (b) is
with the nearest-neighbor interlayer correlation (not depicted). We find
the latter to be AFM in phase I and FM in phase II, in agreement with the
Lanczos results. Panel (c) of Fig. \ref{fig:SSCF_zigzag} clearly shows,
that the IDP phase only has short range spin-spin correlations,
consistently with a finite gap.

\begin{figure}[t]
\includegraphics[width=0.45\textwidth]{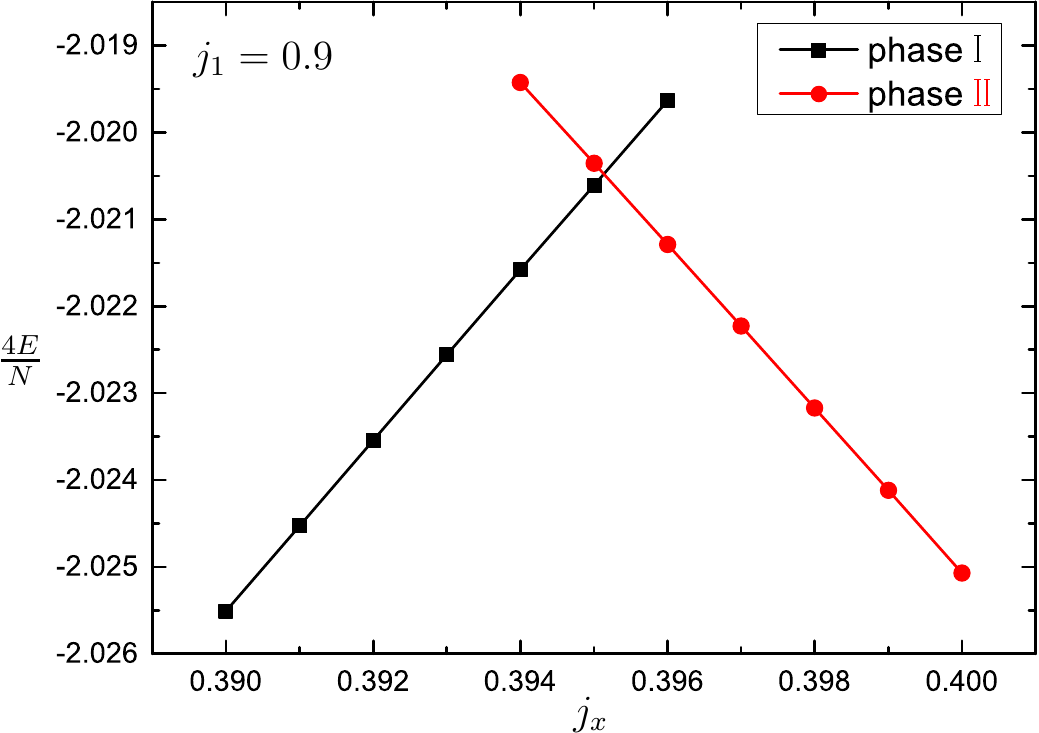}
\caption{(Color online) Energy per unit cell from SB-MFT along a vertical cut through the phase diagram
(Fig.~\ref{fig:phase-diagram}) at $j_{1}=0.9$ for the phase transition
I-II.}
\label{fig:first_order_transition}
\end{figure}

To determine the location of the transitions between the LRO phases we may
use, that these phases have no subgroup relations, and therefore any direct
transitions between them is of first order, i.e. they can be determined
from a discontinuity in the ground state energy. This is true, both, for ED
and SB-MFT. In Fig. \ref{fig:first_order_transition} a representative
example obtained from the latter is depicted for a vertical cut through
Fig. \ref{fig:phase-diagram}. Similar results are obtained from ED and will
be summarized in the next Section.

\section{Quantum phase diagram}
\label{sec:Quantum-phases}

\begin{figure}[t]
\includegraphics[width=0.4\textwidth]{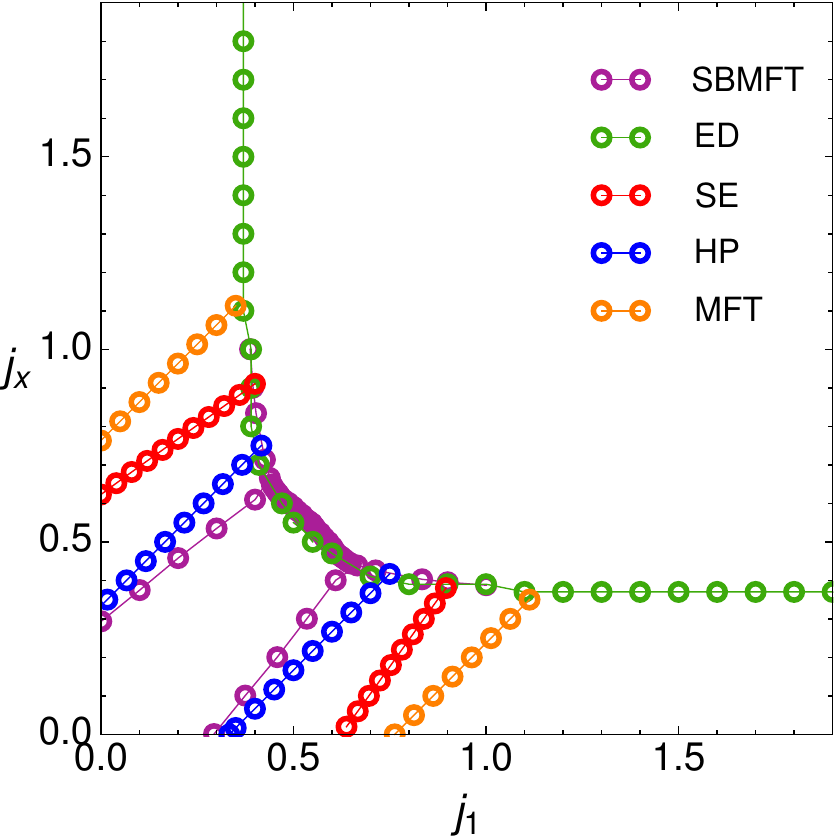}
\caption{Quantum phases and critical lines determined by the different
techniques considered.}
\label{fig:SW-phases}
\vskip-.3cm
\end{figure}

In this Section we compare the critical lines for the phase transitions of
the system obtained from all complementary methods of this work. As a
central result Fig. \ref{fig:SW-phases} compiles our findings from SB-MFT,
BO-HP, BO-MFT, D-SE, and ED. This figure is the {\em quantitative} analog of
the {\em qualitative} sketch in Fig. \ref{fig:phase-diagram}. Several comments
are in order.

To begin, we note that for the first order transitions, i.e.
I$\leftrightarrow$II, II$\leftrightarrow$III, and IDP$\leftrightarrow$II,
there is a very good quantitative agreement between SB-MFT and ED, showed
by line-connected magenta and green open circles, respectively in
Fig. \ref{fig:SW-phases}. This is expected, since first order transitions
are determined by ground state energies. These are less susceptible to
errors of different approaches as e.g. finite size effects or mean-field
approximations. We note that SB-MFT technique is the only method employed
in our work, which potentially allows for an estimation of all critical
lines, independently of the character of the transition, i.e. first or
second order.

In contrast to the first order transitions, for the second order
IDP$\leftrightarrow$(I, III) transitions, the critical lines obtained from
our complementary methods will determine a range of potential transition
points at most, since the gap closure, i.e. the behavior of the critical
correlation length is sensitive to the method used. Nevertheless it is
clearly visible from Fig. \ref{fig:SW-phases}, that the symmetric regions
of both IDP$\leftrightarrow$(I, III) transitions are centered around the
lines $j_x\sim j_1 \pm 0.6(\pm 0.2)$, where $\pm 0.2$ denotes an
uncertainty set by the scatter between the various approaches. Note that
this scatter also implies an uncertainty of the location of the two
tricritical points separating phases IDP-I-II and IDP-II-III.

Remarkably all techniques predict essentially straight critical lines for
the IDP$\leftrightarrow$(I, III) transitions with approximately unit slope,
at least on the scale of the plot. This is a direct consequence of the last
term in Eq. \ref{eq:H0}, perturbing the exact dimer state. As a
consequence, e.g. in both BO methods, and by construction, the triplon
hopping amplitude is a function of the combination of exchanges
$|j_{1}-j_{x}|$ only. Yet, D-SE at $O(4)$ (red open circles in
Fig. \ref{fig:SW-phases}) leads to a small curvature of the transition
lines. In BO-HP it is possible to obtain an analytical expression, namely
$j_{x} = j_{1} \pm 1/3$, for critical lines (see appendix
\ref{sec:app-BOMFT}), depicted by blue open circles in
Fig. \ref{fig:SW-phases}. For BO-MFT (orange open circles in
Fig. \ref{fig:SW-phases}), the offset $1/3$ is replaced through numerical
solution of the analytic self-consistency equations by $\approx 0.76$ (see
Fig. \ref{fig:gap-bilayer}a))

Note that in all the cases (except SB-MFT) the second order critical line
ends at the border of phase II, which is obviously an artifact of the
methods since, as we have previously mentioned, level crossings are
not detected by D-SE nor BO techniques.

\section{Intralayer frustration}
\label{sec:J2}

\begin{figure}[tb]
\includegraphics[width=0.95\columnwidth]{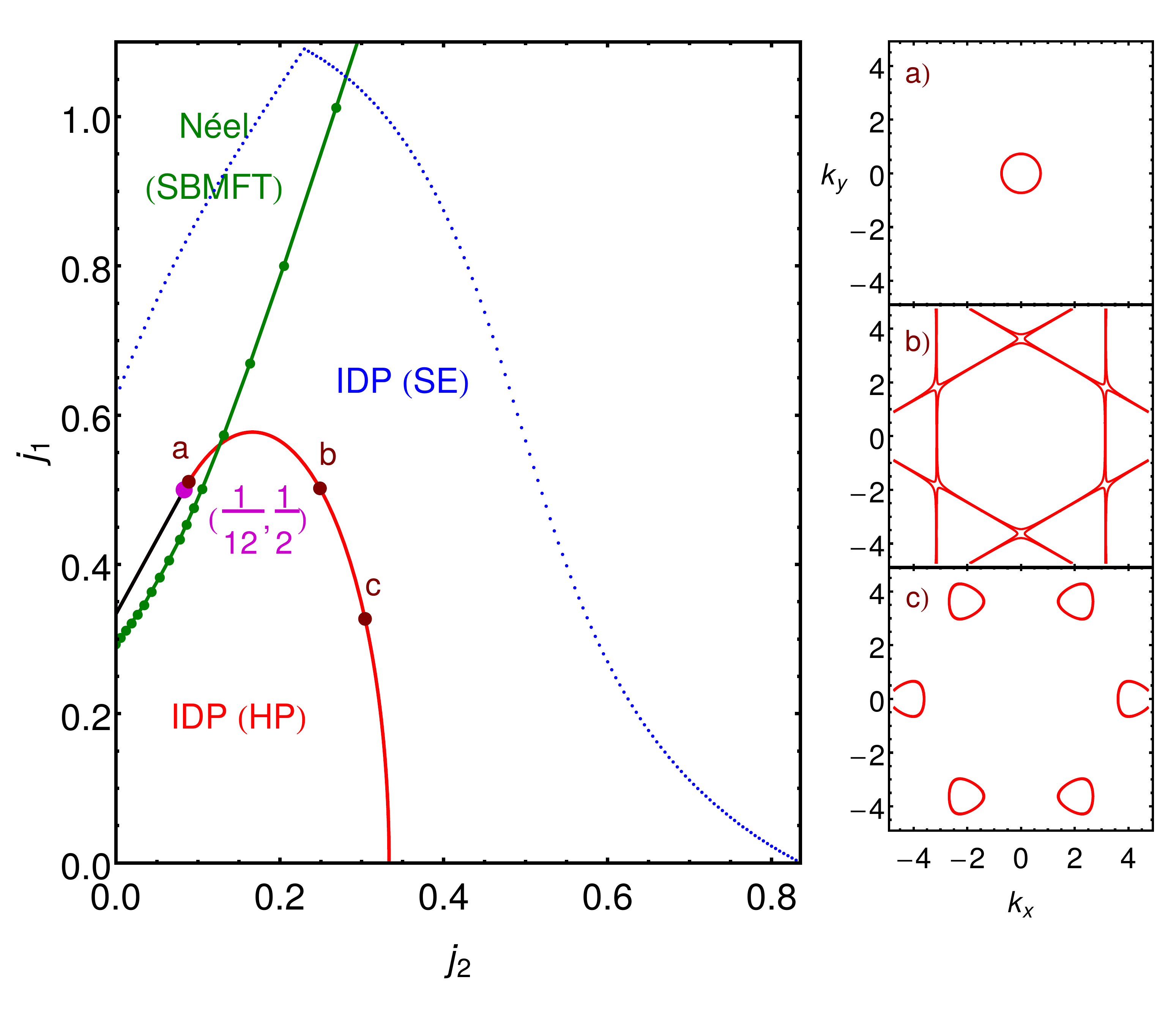} \caption{Left panel:
quantum phases versus $j_1$ and $j_2$ at $j_x=0$ from SBMFT (dotted
green), D-SE (blue dots), as well as BO-HP (black and red). Magenta point
refers to termination point of $k_c=(0,0)$ transitions corresponding to
black line. Right panel: location of critical wave vector for spin gap
closure from BO-HP on red line. Labels a)$-$c) correspond to points on the
critical line.}
\label{fig:j1j2phases}
\end{figure}

In this Section we digress to make contact with previous analysis of the
Heisenberg model on the hexagonal bilayer including a next-nearest neighbor
frustrating {\it intralayer} exchange $j_2$ \cite{Zhang2014}. In the latter
work, the N\'eel phase (identified as phase I in the present work), which
is known to exist in the single layer model for $j_2/j_1\lesssim 0.2$
\cite{Bishop_2012,Mezzacapo,Zhang_PRB_2013} was shown to persist within a
finite region of interlayer coupling, including a re-entrant pocket at
small $j_0/j_1$. Beyond the N\'eel phase a quantum disordered region was
predicted (see Fig. 3 of ref. \onlinecite{Zhang2014}). Here we clarify to
which extend this disordered phase is connected to the IDP discussed in
Section \ref{sec:IDP}. To this end Fig. \ref{fig:j1j2phases} shows the
quantum critical lines of the gap closure, both from BO-HP and D-SE versus
$j_{1,2}$ at $j_x=0$, combined with those from the SBMFT from
ref. \onlinecite{Zhang_PRB_2013}. Note that for this comparison the
dimensionless exchange parameters used in the latter reference,
i.e. $J_2/J_1$, $J_0/J_1$, need to be rescaled onto $j_1=J_1/J_0$,
$j_2=J_2/J_0$, used in the present manuscript.

Several comments are in order. First, we note that the intersections of the
results from all three methods on the $j_1$-axis necessarily are {\it identical}
to those occurring on the same axis in Fig. \ref{fig:SW-phases}. Next, we
focus on BO-HP versus SB-MFT. As is obvious, the IDP emerges as a new phase
in the $j_1$-$j_2$ plane, which had not been identified in ref.
\onlinecite{Zhang_PRB_2013}, into which the N\'eel phase (I) transforms. The
corresponding quantum critical line from the BO-HP is dissected into a
black line-segment, which terminates at the point $(j_2,j_1)=(1/12,1/2)$
(magenta) and a red line-segment. On the black line-segment the critical
wave vector is $k_c=(0,0)$. This is consistent with the transition into the
N\'eel state, obtained from the SB-MFT approach (indicated by green dots
in Fig. \ref{fig:j1j2phases}). Interestingly, not only the
critical point directly at $j_2=0$ is very similar between BO-HP and SBMFT,
but all of the critical boundaries nearly coincide up to the magenta point,
where $k_c$ from BO-HP starts to be inconsistent with a transition into a
N\'eel state. This may imply a tricritical point in this range, where
however the nature of the third phase, appearing in the right region of the
left panel of Fig. \ref{fig:j1j2phases} remains unclear. Regarding the
transitions along the red BO-HP line, triplon {\em degeneracy} occurs. This
is exemplified for three selected points: a, b, and c on the critical line,
for which the right panel displays the locations of the gap-closure in
$k$-space. Apart from the appearance of degenerate line zeros in the latter
panel, Fig. \ref{fig:j1j2phases}a)$-$c) clearly show the evolution from a
N\'eel state at $j_2\ll 1$ and sufficiently large $j_1$ into the
120\degree-order of the triangular lattice antiferromagnet, into which the
bilayer decomposes for $j_1=0$ and $j_2$ above the critical value of 1/2.

Line-zeros of the triplon dispersion on the quadratic level of the BO
theory have dramatic consequences for BO-MFT. Power counting for
Eq. (\ref{wa13}) shows that the integral on the right hand side diverges
for such cases. In turn the MFT gap stays {\em finite} for all $j_2/j_1$
displaying line {\em minima} of the dispersion. This renders BO-MFT
applicable only for transitions into commensurate phases of the hexagonal
bilayer lattice. While, to the best of our knowledge, such behavior has not been
reported for BO-MFT on other dimer quantum magnets, it is certainly an
artifact of the quadratic approximation and requires future analysis. Here
we refrain from using BO-MFT with intralayer frustration.

Regarding the D-SE results, Fig. \ref{fig:j1j2phases} shows that the region
of stability for the IDP in the $j_1$-$j_2$ plane turns out to be
significantly larger than for the BO-HP, which is similar to the
conclusions drawn in the $j_1$-$j_x$ plane. In contrast to the quadratic
approximation, the spin gap in $O(4)$ D-SE closes at the single point
$k_c=(0,0)$ for the range $0 \leq j_2 \lesssim 0.8$. However, around the
point $j_2 \approx 0.8$ (and $j_1 \approx 0$) other modes start to compete.
It would be necessary to go to higher orders in the series to clarify the
k-dependence of gap closure for such a large values of $j_2$. This issue is
beyond the scope of present analysis.

\section{Conclusions}
\label{sec:conclusions}

We have studied the zero temperature quantum phase diagram of the
frustrated antiferromagnet on the bilayer honeycomb lattice. To
characterize the different phases present in the model, as well as their
transitions, we have calculated a variety of quantities, such as ground
state energies, low energy excitations, triplet gaps and static spin-spin
correlations. This has been done, using several complementary techniques:
bond operator and Schwinger bosons mean field theories, dimer series
expansion and exact diagonalization of finite systems.

The main results of our work are contained in the schematic phase diagram
of Fig. \ref{fig:phase-diagram}. This diagram is symmetric with respect to
$j_{1}=j_{x}$. For $j_{1}=j_{x} \leq j_x^c \approx 0.55$ the model exhibits
an exact interlayer dimer-product state, whose ground state and elementary
triplet excitations are identical to the decoupled dimer limit
($j_{1}=j_{x}=0$). Perpendicularly to the diagonal line a dimerised phase
evolves adiabatically from the exact ground state and extends over a region
around the diagonal line. This gapped \emph{interlayer dimer phase} (IDP)
has been analyzed by means of bond operator theory and dimer series
expansion (complemented with Lanczos diagonalization) since both methods
are exact for the singlet product state.

In contrast to the IDP phase, which is a gapped, magnetically disordered,
and of quantum origin, the other phases present in the model are gapless,
magnetically ordered, and quasi classical. In particular we have determined
three magnetic phases, denoted by I, II, and III in
Fig. \ref{fig:phase-diagram}. The phases I and II are N\'eel-like, whereas
III exhibits columnar order. The magnetic structure of these phases has
been clarified both, by exact diagonalization on finite systems of N=24
sites and by Schwinger bosons mean field theory on large lattices of
N=10000 sites, both with identical results. In particular phase II along
the diagonal line, for $j_{1}=j_{x} > j_x^c$ is equivalent to the ground
state of an effective spin-1 Heisenberg model on the single-layer honeycomb
lattice, with an antiferromagnetic coupling $j_1=j_x$.

The nature of the phase transitions has been identified as first order
(level crossing) for the transitions I$\leftrightarrow$II,
II$\leftrightarrow$III and IDP$\leftrightarrow$II, and second order (gap
closure) for the transitions IDP$\leftrightarrow$I and
IDP$\leftrightarrow$III. A quantitative analysis of the quantum phase
diagram, obtained from the combination of all methods has been
presented. For all first order transitions good agreement between Lanczos
and Schwinger bosons MFT has been obtained. For the second order
transitions, qualitative agreement between the different methods used has
been shown.

Finally we have briefly explored the effects of intralayer frustration. We
find, that both, the IDP and the LRO phase I naturally extend into the
$j_1{-}j_2$ plane, and are terminated by sufficiently large intralayer
frustration $j_2$.

\begin{acknowledgments}
H.Z. thanks Lu Yu for fruitful discussions, and the Institute of Physics, Chinese 
Academy of Sciences for financial support. C.A.L. and M.A. are supported by CONICET and FONCyT (PICT 2013-0009). 
Work of W.B. has been supported in part by the Deutsche Forschungsgemeinschaft
through SFB 1143 and by the National Science Foundation under Grant No. NSF
PHY11-25915. W.B. also acknowledges kind hospitality of the Platform for
Superconductivity and Magnetism, Dresden.
\end{acknowledgments}


\appendix

\section{Bond Operator Approach}
\label{sec:app-BOMFT}

Quantum spin models comprising weakly coupled antiferromagnetic spin-1/2
dimers allow for a description in terms of bosonic operators, so called
\emph{bond operators} (BO) \cite{Chubukov1989,Chubukov1991,Sachdev1990},
which label the dimer's singlet-triplet spectrum. BOs lead to a treatment
of dimerised phases similar to the linear spin wave theory for magnetically
ordered phases. Within BO theory the two spins $\vec{S}_{i=1,2}$ on each
dimer are expressed as
\begin{equation}
S_{\stackrel{{\scriptstyle 1}}{{\scriptstyle 2}}}^{\alpha}=\frac{1}{2}(\pm
s^{\dagger}t_{\alpha}\pm t^{\dagger}s-\sum_{\beta,\gamma}i
\varepsilon_{\alpha\beta\gamma}t_{\beta}^{\dagger}t_{\gamma}^{\phantom{\dagger}})
\,,\label{wa1}
\end{equation}
where $s^{(\dagger)}$and $t_{\alpha}^{(\dagger)}$ destroy(create)
the singlet and triplet states of the dimer and Greek labels, $\alpha=1,2,3$,
refer to the threefold triplet multiplet. A hard-core constraint
\begin{equation}
s^{\dagger}s+\sum_{\alpha}t_{\alpha}^{\dagger}
t_{\alpha}^{\phantom{\dagger}}=1\label{wa2}
\end{equation}
is implied, which renders the algebra of the r.h.s of Eq. (\ref{wa1})
identical to that of spins.

Inserting the BO representation into a spin model leads to an interacting
Bose gas. Two kinds of \emph{quadratic approximations} have become popular
in the limit of weak dimer coupling, namely the BO mean-field theory (BO-MFT)
\cite{Sachdev1990} and the BO Holstein-Primakoff (BO-HP) approach
\cite{Chubukov1989,Chubukov1991}. In both cases, terms only up to second
order in the BOs are retained. In the BO-MFT, singlets are condensed by
$s^{(\dagger)}\rightarrow s\in\,$Re and the constraint Eq. (\ref{wa2}) is
satisfied on the average with a global Lagrange multiplier $\eta$
\cite{Sachdev1990}. In the BO-HP, the constraint is used to eliminate all
singlet operators using $s^{\phantom{\dagger}}= s^{\dagger}=
(1-\sum_{\alpha}t_{\alpha}^{\dagger} t_{\alpha}^{\phantom{\dagger}}
)^{-1/2}$, followed by expanding the square root
\cite{Chubukov1989,Chubukov1991}.

For both approaches, i.e. BO-MFT and BO-HP, the Hamiltonian in units
of $J_{0}$ of our frustrated hexagonal bilayer lattice reads
\begin{eqnarray}
H & = & H_{0}+H_{1}+H_{2}+H_{c}\label{wa3}\\
H_{0} & = & \sum_{l,b}(-\frac{3}{4}s^{2}+\frac{1}{2}
\sum_{\alpha}t_{lb\alpha}^{\dagger}
t_{lb\alpha}^{\phantom{\dagger}})
\label{wa4}\\
H_{1} & = & \sum_{l,\widetilde{m},\alpha}
\frac{s^{2}\widetilde{j_{1}}}{2}(t_{\widetilde{m}
A\alpha}^{\dagger}t_{lB\alpha}^{\phantom{\dagger}}+
t_{\widetilde{m}A\alpha}^{\dagger}t_{lB\alpha}^{\dagger}+h.c.)
\label{wa5}\\
H_{2} & = & \sum_{l,\widetilde{l},\alpha,b}
\frac{s^{2}j_{2}}{2}(t_{\widetilde{l}b\alpha}^{\dagger}
t_{lb\alpha}^{\phantom{\dagger}}+t_{\widetilde{l}b\alpha}^{\dagger}
t_{lb\alpha}^{\dagger}+h.c.)
\label{wa6}\\
H_{c} & = & -\sum_{l,b}\eta(s^{2}+\sum_{\alpha}
t_{lb\alpha}^{\dagger}t_{lb\alpha}^{\phantom{\dagger}}-1)
\label{wa7}
\end{eqnarray}
where $t_{lb\alpha}^{(\dagger)}$ labels triplets in unit cell $l$ at basis
site $b=A,B$ of the two interpenetrating triangular lattices comprising the
hexagonal lattice. The sites $\widetilde{m}A$ in Eq.  (\ref{wa5}) refer to
the three nearest neighbors of the honeycomb basis around each of the
triangular lattice sites at $lB$ and the $\widetilde{l}$ labels the three
nearest neighbors on each of the triangular lattices. $\widetilde{j}_{1}=
j_1-j_x$ and $j_{2}$ are the dimensionless exchange couplings.  $s^{2}$ is the singlet condensate, and
$\eta$ the global Lagrange multiplier for constraint (\ref{wa2}).

This Hamiltonian can be diagonalized by standard Bogoliubov transformation
leading to an energy $E$ \emph{per unit cell}, i.e. per two dimers, of
\begin{equation}
E={-}\frac{3}{4}{-}\frac{3}{2}s^{2}{-}2\eta s^{2}{+}
5\eta+\frac{3}{2N}\sum_{k}(E_{k+}{+}E_{k-})\label{wa8}
\end{equation}
with the \emph{triplon dispersion}
\begin{equation}
E_{k\pm}=a\sqrt{1\pm\frac{s^{2}}{a}e_{\pm}(k)}\label{wa9}
\end{equation}
where
\begin{eqnarray}
e_{\pm}(k) & =\widetilde{j}_{1} & \sqrt{3{+}2\cos(k_{x}){+}
4\cos(\frac{k_{x}}{2})\cos(\frac{\sqrt{3}k_{y}}{2})}
\hphantom{aaaa}\nonumber \\
 &  & \pm2j_{2}(\cos(k_{x}){+}2\cos(\frac{k_{x}}{2})
\cos(\frac{\sqrt{3}k_{y}}{2}))
\label{wa10}\\
 & \equiv\widetilde{j}_{1} & \sqrt{3+g(k)}\pm j_{2}\,g(k)
\label{wa11}
\end{eqnarray}
and $a=1/4-\eta$. Eqs. (\ref{wa9})-(\ref{wa11}) display an important
symmetry for $\widetilde{j}_{1}\leftrightarrow-\widetilde{j}_{1}$, namely
for that $e_{\pm}(k)\leftrightarrow-e_{\mp}(k)$. This implies, that on the
quadratic level of the BO-HP and BO-MFT all results of the theory will be
symmetric w.r.t. diagonal $j_{1}=j_{x}$

From (\ref{wa8})-(\ref{wa11}) the BO-HP is completed by replacing the sum
of the first four addends in Eq. (\ref{wa8}) with to $-9/2$ and by setting
$a=1$, $s=1$ in (\ref{wa9}, \ref{wa10}).

For the BO-MFT the energy $E$ has to be extremized, implying two
selfconsistency equations $\partial_{a}E/\partial a=0$ and $\partial_{s}E /
\partial s=0$.  These can be combined into a single one for the parameter
$d=s^{2}/a$, i.e.
\begin{equation}
d=\frac{5}{2}-\frac{3}{4N}\sum_{k,v=\pm}\frac{1}{\sqrt{1+v\,d\,e_{v}(k)}}
\,.\label{wa12}
\end{equation}
Knowing $d$, both mean field parameters can be obtained from substituting
into either one of the mean field equations, e.g. $\partial_{a}E/\partial
a=0$
\begin{equation}
2s^{2}=5-\frac{3}{2N}\sum_{k,v=\pm}\frac{1+\frac{1}{2}\,v\,
d\,e_{v}(k)}{\sqrt{1+v\,d\,e_{v}(k)}}
\,.\label{wa13}
\end{equation}
We mention in passing, that the trivial limit, i.e. $\widetilde{j}_{1}
=j_{2} =0$, leads to $d=1$, $s=1$, and $\eta=-3{/}4$, and therefore to a
\emph{singlet-triplet gap} of $\Delta=1$ and a ground state energy of
$E={-}3{/}2$, which is consistent with two saturated singlets per unit
cell.
\color{black}

\section{Schwinger Boson Mean-Field Approach}
\label{sub:SW-MFT}

In the Schwinger-boson representation, the Heisenberg interaction
can be written as a biquadratic form. The spin operators are
replaced by two species of bosons via the
relation\cite{Auerbach,Auerbach:1994,Auerbach:2011}
 \ba
 \vec{\mathbf{S}}_{\alpha}(\vec{r})=\frac{1}{2}\vec{\mathbf{b}}_{\alpha}^{\dag}(\vec{r})\cdot\vec{\sigma}\cdot\vec{\mathbf{b}}_{\alpha}(\vec{r}),
\ea
 where ${\vec{\bf b}_{\alpha}(\vec{r})^{\dagger }}\!=\!({\bf b}^{\dagger }_{\alpha,\uparrow }(\vec{r}),{\bf b}^{\dagger }_{\alpha,\downarrow }(\vec{r}))$
 is a bosonic spinor corresponding to the site $\alpha$ in the
 unit cell sitting at $\vec{r}$.  $\vec{\sigma}$ is the vector of
Pauli matrices, and there is a boson-number restriction
$\sum_\sigma \mathbf{b}^{\dag}_{\alpha,\sigma}(\vec{r})\mathbf{b}_{\alpha,\sigma}(\vec{r})\!=\!2S$ on each site.

In terms of boson operators we define the $SU(2)$ invariants
\ba
 \mathbf{A}_{\alpha \beta}(\vec{x},\vec{y})&=&\frac12 \sum_{\sigma} \sigma
\mathbf{b}_{\alpha,\sigma}(\vec{x})\mathbf{b}_{\beta,-\sigma}(\vec{y})\\
\mathbf{B}_{\alpha \beta}(\vec{x},\vec{y})&=&\frac12 \sum_{\sigma}
\mathbf{b}^{\dag}_{\alpha,\sigma}(\vec{x})\mathbf{b}_{\beta,-\sigma}(\vec{y}).
\ea
\normalsize
The operator $\mathbf{A}_{\alpha \beta}(\vec{x},\vec{y})$ creates a
spin singlet pair between sites $\alpha$ and  $\beta$ corresponding
to unit cells located at $\vec{x}$ and
 $\vec{y}$ respectively.
The operator $\mathbf{B}_{\alpha \beta}(\vec{x},\vec{y})$ creates a
ferromagnetic bond, which implies the intersite coherent hopping of
the Schwinger bosons.

In this representation, the rotational invariant spin-spin
interaction can be written as
\small
\ba
\nonumber
  \vec{\mathbf{S}}_{\alpha}(\vec{x})\cdot \vec{\mathbf{S}}_{\beta}(\vec{y})
  =:\mathbf{B}^{\dag}_{\alpha \beta}(\vec{x},\vec{y}) \mathbf{B}_{\alpha \beta}(\vec{x},\vec{y}):
   -\mathbf{A}^{\dag}_{\alpha \beta}(\vec{x},\vec{y}) \mathbf{A}_{\alpha \beta}(\vec{x},\vec{y})
\ea
\normalsize
where $:\mathbf{O}:$ denotes the normal ordering of the operator $\mathbf{O}$.
One of the advantages of this rotational invariant decomposition is
that it enables to treat ferromagnetism and antiferromagnetism on
equal footing. This decomposition has been successfully used to
describe quantum disordered phases in two-dimensional frustrated
antiferromagnets\cite{Cabra_honeycomb_prb,Cabra_honeycomb_2,Zhang_PRB_2013,Trumper1,Trumper2,Mezio,Messio,Messio_2013}.

In order to generate a mean field theory, we perform the Hartree-Fock decoupling
\ba
\nonumber
(\vec{\mathbf{S}}_{\alpha}(\vec{x})\cdot \vec{\mathbf{S}}_{\beta}(\vec{y}) )_{MF}&=&
 [B^{*}_{\alpha \beta}(\vec{x}-\vec{y}) \mathbf{B}_{\alpha \beta}(\vec{x},\vec{y})\\
 &-& A^{*}_{\alpha \beta}(\vec{x}-\vec{y}) \mathbf{A}_{\alpha \beta}(\vec{x},\vec{y})]\\\nonumber
&-& \langle (\vec{\mathbf{S}}_{\alpha}(\vec{x})\cdot \vec{\mathbf{S}}_{\beta}(\vec{y}) )_{MF}
\rangle
\ea
where the mean field parameters are given by
\ba
\label{eq:mfeqs}
A^{*}_{\alpha \beta}(\vec{x}-\vec{y})&=&\langle \mathbf{A}^{\dag}_{\alpha \beta}(\vec{x},\vec{y})\rangle,\\
\label{eq:mfeqs2}
B^{*}_{\alpha \beta}(\vec{x}-\vec{y})&=&\langle \mathbf{B}^{\dag}_{\alpha \beta}(\vec{x},\vec{y})\rangle ,
\ea
and the exchange at the mean field level is
\ba
\langle (\vec{\mathbf{S}}_{\alpha}(\vec{x})\cdot \vec{\mathbf{S}}_{\beta}(\vec{y}) )_{MF} \rangle
= |B_{\alpha \beta}(\vec{x}-\vec{y})|^{2}-|A_{\alpha \beta}(\vec{x}-\vec{y})|^{2}.\nn\\
\ea
The mean field equations (\ref{eq:mfeqs}) and (\ref{eq:mfeqs2}) must
be solved in a self-consistent way together with the following
constraint for the number of bosons in the system
\ba \label{eq:constraint} B_{\alpha
\alpha}(\vec{R}=\vec{0})=4N_{c}S, \ea
where $N_c$ is the total number of unit cells and $S$ is the spin
strength.
Self-consistent solutions in the bilayer honeycomb lattice involve
finding the roots of coupled nonlinear equations for the mean field
parameters and solving the constraints to determine the values of
the Lagrange multipliers $\lambda^{(\alpha)}$ which fix the number
of bosons in the system. We perform the calculations for large
systems and extrapolate the results to the thermodynamic limit.
Details of the self consistent calculation can be found in Refs. \onlinecite{Cabra_honeycomb_prb,Zhang_PRB_2013}.

\section{Series expansion\label{sub:Series-expansion}}

The D-SE calculations start from the limit of isolated dimers. To
this end we decompose the Hamiltonian given by Eq.(\ref{eq:Hspin_general}) in units of $J_0$ into \begin{equation}
H=H_0 + V(j_1,j_x,j_2),\label{se1}\end{equation}
where $H_0$ represents decoupled interlayer dimers and $V(j_1,j_x,j_2)$
is the interaction part of Hamiltonian, connecting dimers via $j_1,j_x,j_2$
couplings.

By construction the levels structure of $H_0$ is equidistant, which allows
to sort the spectrum of $H_0$ in a block-diagonal form, where
each block is labeled by an energy quantum-number
Q. Therefore, Q=0 represents the ground state (\emph{vacuum}), i.e.,
all dimers are in the singlet state. Q=1 sector is composed
by states obtained by creating (from vacuum state) one-elementary
triplet excitation (\emph{particle}) on a given dimer, and so on. The cases
in which $Q\geq2$ will be of multiparticle nature.

In general the action of $V(j_1,j_x,j_2)$ mixes different Q-sectors, so
that the block-diagonal form of $H_0$ is not conserved in $H$.  However
because of the ladder structure of the unperturbed spectrum, is possible to
restore the block-diagonal form by application of continuous unitary
transformations, using the flow equation method of Wegner
\cite{Wegner-1994a}. This method can be implemented perturbatively order by
order, transforming $H$ onto an effective Hamiltonian $H_{\mathrm{eff}}$
which is block-diagonal in the quantum number $Q$, having the structure
\begin{equation}
H_{\mathrm{eff}}=H_0 + \sum_{n,m,l}^\infty C_{n,m,l} j_1^{n}j_x^{m}j_{2}^{l},\label{eq:5}\end{equation}
where $C_{n,m,l}$ are weighted products of terms in $V(j_1,j_x,j_2)$
which conserve the $Q$-number, with weights determined by recursive
differential equations, details of which can be found in Ref. \onlinecite{Knetter2000a}.

Due to $Q$-number conservation several observables can be calculated
directly from $H_{\mathrm{eff}}$ in terms of a D-SE in $j_1,j_x,j_2$.  For
systems with coupled spin-plaquettes continuous unitary transformations D-SE
has been used for one \cite{Arlego2006a}, two
\cite{Brenig2002aa,Brenig2004a,Arlego2007a,Arlego2008a} and three
\cite{Brenig2003a} dimensions. For the present model we have performed
$O(4)$ D-SE in $j_1,j_x,j_2$ for ground state energy ($Q=0$) and for $Q=1$
sectors, respectively. We refer for technical details about the calculation
to ref. \cite{Arlego2011}.  Note finally that the contribution of
perturbation in the case $V(j_1,j_1,0)$ is \emph{zero}, reflecting the
invariance of original Hamiltonian under $j_1 \leftrightarrow j_x$ exchange.

\section{Linear Spin Wave Theory at $j_{1}=j_x$
\label{sec:Linear-Spin-Wave}}

Here we briefly quote the equations necessary to determine the critical
coupling $j_{1}^c$ for the first order IDP$\leftrightarrow$II quantum phase
transition along the line $j_{1}=j_{x}$ from linear spin wave
theory. In the IDP along the latter line, the ground state energy is
\begin{equation}
E_{\textrm{all \ensuremath{L=0} sector}}/J_{0}=-\frac{3}{2}N_{\triangle}
\,,\label{eq:B1}
\end{equation}
where $N_{\triangle}$ is the number of triangular unit cells. The
Hamiltonian of the ``all $L=1$ sector'' on the other hand reads
\begin{equation}
H_{\textrm{all \ensuremath{L=1} sector}}/J_{0}=\frac{1}{2}N_{\triangle}+
j_{1}\sum_{\langle lm\rangle}\mathbf{L}_{l}\cdot\mathbf{L}_{m}
\,,\label{eq:B2}
\end{equation}
where the sum refers to an $S=1$ Heisenberg antiferromagnet on the
hexagonal lattice. The ground state of the latter is known to be an N\'eel
state with an energy per \emph{site} to $O(1/S)$ of \cite{Rastelli1979}
\begin{eqnarray}
\lefteqn{E_{LSWT}=j_{1}\left\{ -\frac{3S^{2}}{2}+\frac{S}{4\pi^{2}\sqrt{2}}
\int_{0}^{2\pi}\int_{0}^{2\pi}dx\,dy\,[3{-}\right.}
\nonumber \\
 &  & \hphantom{aaaaaaa}\left.\cos(x){-}\cos(y){-}\cos(x{+}y)]^{1/2}{-}
\frac{3S}{2}\right\} \hphantom{aaa}
\nonumber \\
 & \simeq & j_{1}\left(-\frac{3S^{2}}{2}-0.314763\,S\right)
\,.\label{eq:B3}
\end{eqnarray}
For $S=1$ this yields the line
\begin{equation}
E_{LSWT}\simeq-1.81476\,j_{1}\,,\label{eq:B4}
\end{equation}
which is plotted in Fig. \ref{fig:energy-exact}. Together with (\ref{eq:B1},
\ref{eq:B2}) and keeping in mind that a '\emph{site}' in (\ref{eq:B3})
refers to \emph{two} spins on the original bilayer, this implies that
$1.81476\,j_{1}^c=1$, i.e.
\[
j_{1}^c\simeq0.551036
\]

\vspace{1cm}

\bibliography{allrefs}
\end{document}